\documentclass[aip,rsi,reprint,graphicx]{revtex4-1}
\usepackage{graphicx}
\usepackage{amsmath}

\newcommand{\ket}[1]{|{#1}\rangle}

\newcommand{\be}{\begin{equation}}
\newcommand{\ee}{\end{equation}}

\begin{document}
\title{In-situ magnetometry for experiments with atomic quantum gases}

\author{Ludwig Krinner, Michael Stewart, Arturo Pazmi\~no, and Dominik Schneble}

\affiliation{Department of Physics and Astronomy, Stony Brook University, Stony Brook, New York 11794-3800, USA}

\date{\today}

\begin{abstract}
Precise control of magnetic fields is a frequent challenge encountered in experiments with atomic quantum gases. Here we present a simple method for performing in-situ monitoring of magnetic fields that can readily be implemented in any quantum-gas apparatus in which a dedicated field-stabilization approach is not possible. The method, which works by sampling several Rabi resonances between magnetically field sensitive internal states that are not otherwise used in a given experiment, can be integrated with standard measurement sequences at arbitrary fields.  For a condensate of $^{87}$Rb atoms, we demonstrate the reconstruction of Gauss-level bias fields with an accuracy of tens of microgauss and with millisecond time resolution. We test the performance of the method using measurements of slow resonant Rabi oscillations on a magnetic-field sensitive transition, and give an example for its use in experiments with state-selective optical potentials. 
\end{abstract}

\maketitle

\section{Introduction}

Experiments with ultracold atomic quantum gases \cite{RN28382} often call for the manipulation and control of the atoms' spin degree of freedom, including work with spinor condensates~\cite{RN12416} or homonuclear atomic mixtures in state-selective optical potentials \cite{RN3002,RN2911,RN2912,RN3203,RN14225,RN19828,RN17365,RN3055} where a control of Zeeman energies to a fraction of the chemical potential (typically on the order of one kilohertz or one milligauss), may be required.  With fluctuations and slow drifts of ambient laboratory magnetic fields on the order of several to tens of milligauss, achieving such a degree of control over an extended amount of time requires dedicated field-stabilization techniques. However, in a multi-purpose BEC machine, this may be challenging given geometric constraints that can interfere with shielding or with placing magnetic-field probes sufficiently close to an atomic cloud, which are often subject to short-range, drifting stray fields from nearby vacuum hardware or optomechanical mounts. To address this problem, we have developed a simple method for direct monitoring of the  magnetic field at the exact position of the atomic cloud, by employing the cloud as its own field probe, in a way that does not interfere with its originally intended use. The idea is that hyperfine ground-state Zeeman sublevels that are not used in an experimental run can be employed for a rapid, concurrent sampling of Rabi resonances, in the same run, thus making it possible to record and ``tag on'' field information to standard absorption images, which can be used both for slow feedback control or for stable-field postselection. We emphasize that our pulsed, single-shot method, which features an accuracy of tens of microgauss and has an effective bandwidth of one kilohertz, is not meant to compete with state-of-the-art atomic magnetometers \cite{RN28384,RN28385,RN28387,RN2693,RN28386,RN4624}; rather, its distinguishing feature is that it can be implemented without additional hardware and independently of geometric constraints, while featuring a performance that is competitive with that of advanced techniques for field stabilization in a dedicated apparatus\cite{RN2804,RN4429}. It can, at least in principle, be used over a wide range of magnetic fields, starting in the tens of milligauss range.

This paper is structured as follows. Section II presents the principle and implementation of our method. Section III discusses the expected measurement accuracy as well as an experimental test based on a tagged measurement of slow Rabi oscillations on a magnetic-field sensitive transition. Section IV describes an application featuring the precise characterization of a state-selective optical lattice potential via microwave spectroscopy\cite{RN19828}.

\section{Method and Implementation}

\subsection{Principle of Operation}
\begin{figure}[h!]
\centering
    \includegraphics[width=1\columnwidth]{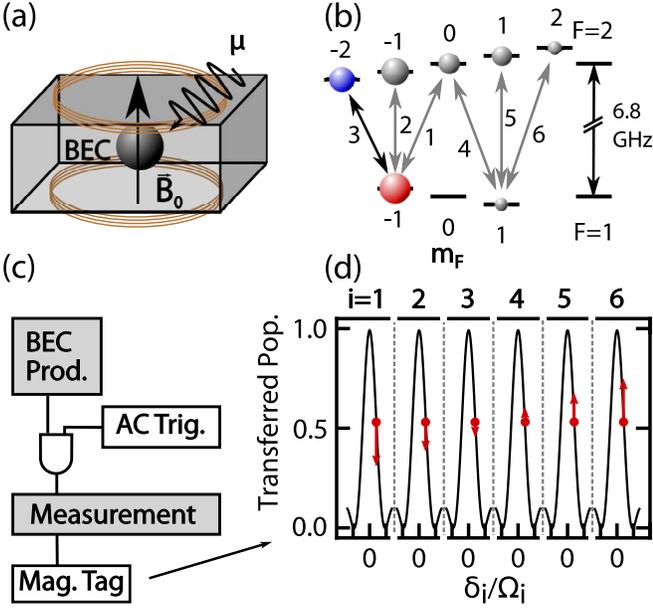}
    \caption{Magnetometry scheme. (a) A Bose-Einstein condensate of $^{87}$Rb atoms in a bias field $B_0$ is subjected to a series of microwave pulses that distributes population over the  $\ket{F,m_F}$ ground state manifold, depending on the exact value of the field. (b) Relevant states for the 6-pulse sequence, the $\ket{1,-1}$ (red) and $\ket{2,-2}$ (blue) states are used for the measurement of Fig. \ref{FIG:RabiCycl}. (c) Outline of a typical experimental run (gray) with the magnetic field tagging added in (white). (d) Rabi resonance for $\Omega_i\tau_i = 0.94\pi$, choice of detunings $\delta_{i}=0.82\Omega_i$ (at $B_0$), and effects of  magnetic-field changes on the transfer probabilities $p_i$ (for identical $\Omega_i$) from which the field is then reconstructed.}
    \label{FIG:Rb87levels}
\end{figure}

The principle of the method is illustrated in Fig.~\ref{FIG:Rb87levels} for the $S_{1/2}(F=1,2)$ hyperfine ground states of $^{87}$Rb, which are split by 6.8 GHz. The atomic sample is located in an externally applied bias field $B_0$ along $z$ leading to a differential Zeeman shift $\delta_z/2\pi =0.7~\mbox{MHz/G}\times B_0$ between neighboring $\ket{F, m_F}$ states. Starting with all atoms in the state $\ket{a}\equiv\ket{F=1,m_F=-1}$, a sequence of microwave pulses $i$ distributes population to $\ket{2,0}, \ket{2,-1}, \ket{2,-2}$ ($i$=1,2,3), and then via $\ket{2,0}$ to $\ket{1,1}$ ($i=4$) and further to $\ket{2,1}$ and $\ket{2,2}$ ($i=5,6$). To ensure isolated addressing of each transition, the detunings $\delta_{i}$ and Rabi couplings $\Omega_i$ are chosen to be small compared to $\delta_z$ (by three orders of magnitude in the example discussed below), and the ordering of the individual pulses is chosen to avoid spurious addressing of near degenerate single photon transitions: $\ket{2,-1}\leftrightarrow \ket{1,0}\approx\ket{2,0}\leftrightarrow \ket{1,-1}$ and $\ket{2,1}\leftrightarrow \ket{1,0}\approx\ket{2,0}\leftrightarrow \ket{1,1}$. Other transitions are near degenerate, but magnetic dipole forbidden, $\left|\Delta m_F\right|>1$. 

The pulse parameters are adjusted such that the final populations $P_{F=2,m_F}$ expected at $B_0$ are comparable, and that their sensitivity to small deviations \footnote{For $\delta B \ll B_0$, fluctuations perpendicular to z can be neglected, since they are quadratically suppressed.} $\delta B$ from $B_0$ is maximal, cf. Fig. \ref{FIG:Rb87levels} (d). Assuming all $\delta_{i}>0$ at $B_0$, the populations change away from the nominal field $B_0$ is negative for $i=1,2,3$ and positive for $i=4,5,6$, with each transition shifted by a different amount. The change in the set of final populations then allows for an unequivocal and precise reconstruction of $B=B_0+\delta B$. In quantitative terms, the transition probabilities $p_i$ for the individual pulses can be calculated from the relative final-state populations $P_{F,m_F}=N(F,m_F)/N$ as
\begin{eqnarray*}
p_1 &=& P_{1,1} + P_{2,0}+ P_{2,1}+P_{2,2}\\
p_2 & =& P_{2,-1}/(P_{1,-1} + P_{2,-1} + P_{2,-2})\\
p_3 &=& P_{2,-2}/(P_{1,-1} + P_{2,-2}) \\
p_4 &=& (P_{1,1}+P_{2,1}+P_{2,2})/(P_{1,1}+P_{2,0}+P_{2,1}+P_{2,2})\\
p_5 &=& P_{2,1}/(P_{1,1}+P_{2,1}+P_{2,2})\\
p_6 &=& P_{2,2}/(P_{1,1}+P_{2,2})
\end{eqnarray*}
Each $p_i$ is related to the magnitude of the magnetic field $B$ via
\begin{equation}
\label{EQ:RabiEq}
p_i = (\Omega_i/\tilde{\Omega}_i)^2 \sin^2(\Omega_i^\prime\tau_i/2),
\end{equation}
where $\tilde{\Omega}_i = (\Omega_i^2+ \delta_i^2)^{1/2}$ and $\delta_i=\delta_i(B)$ is the modified detuning of the $i$th pulse from the $i$th addressed resonance. Assuming that the Rabi couplings $\Omega_i$ are known from an independent calibration, the magnetic field $B$ can then be extracted by fitting $\hbar(\delta_i+\omega_i) = E(F_i, m_{F_i};B)-E(F^\prime_i, m_{F^\prime_i};B)$, where $\omega_i$ is the microwave frequency for the $i$th pulse, and where
\begin{equation}
E(F,m_F;B)=-\frac{\hbar\Delta}{8}\pm\frac{\hbar\Delta}{2} \sqrt{1 + m_Fx + x^2} + g_I\mu_B m_F B
\end{equation}
is the Breit-Rabi energy of the levels involved in the transition, where the +(-) sign holds for F=2(1), $x = (g_I-g_s)\mu_B B/\Delta$, with $g_s$ the g-factor of the electron, $\Delta  = 2 \pi \times 6.834...$ GHz and $g_I = -9.951...\times10^{-4}$ for $^{87}$Rb\cite{RN3163}.

\subsection{Experimental implementation}
\label{ExpImp}
Our experiments are performed in a magnetic transporter apparatus \cite{Pertot-09-Machine},  with an optically trapped condensate of $N\sim1\times10^5$ atoms in the $\ket{a}\equiv\ket{1,-1}$ ground state. At the end of an experimental run (which usually contains steps for the manipulation of the motional and/or internal state of the atoms), the atoms are released,  given about 1 ms to expand (to avoid interaction effects), then subjected to the magnetometry pulse sequence described above, and subsequently detected using absorption imaging. For the determination of the state populations $P_{F,m_F}$ we use Stern-Gerlach separation. In addition, to distinguish the $F=1,2$ states with $|m_F|=1$ (note that the $g_F$ factors in $^{87}$Rb have the same magnitude), absorption imaging of the $F=2$ states is first performed using resonant $F=2\to F^\prime=3$ light, which disperses the F=2 atoms while the $F=1$ atoms continue their free fall. After optical pumping of the $F=1$ atoms to $F=2$ (using $F=1\rightarrow F^{\prime}=2$ light) these atoms are then imaged as well.

Several considerations determine the optimum choice of parameters for the magnetometry pulse sequence.  Maximizing the magnetic-field sensitivity of the $p_i$ (see Eq. \ref{EQ:RabiEq}) for a fixed coupling $\Omega_i$ yields optimum detunings $\delta_{i}\approx 0.58\Omega_i$ (at $B_0$) and pulse durations $\tau_i\approx 1.24\pi\Omega_i^{-1}$ (the pulse area should be kept below $3\pi/2$ in order to avoid sidelobes as high as the main lobe in the Rabi spectrum). Additional minimization of the sensitivity to possible fluctuations of $\Omega_i$ (with microwave amplifiers typically specified only to within 1 dB) modifies these conditions to $\delta_{i}\approx 0.82\Omega_i$ and $\tau_i\approx 0.94\pi\Omega_i^{-1}$, respectively. Ideally, the chosen coupling strength $\Omega_i$ of each transition should be proportional to its gyromagnetic ratio $\gamma_i$. Furthermore, the expected range $\delta B$ of fluctuations around $B_0$ sets the optimum choice of $\Omega_i$ through $\delta B\sim\hbar\Omega_i/\mu_B$, and in turn the accuracy of the measurement goes down with increasing $\Omega_i$. In our experiment, we can comfortably realize kHz-range microwave couplings on all transitions (which are independently calibrated from sampling single Rabi resonances).

\begin{figure}[h!]
\centering
    \includegraphics[width=1\columnwidth]{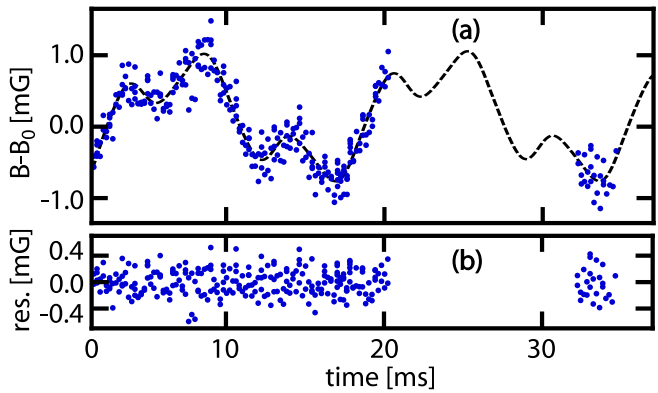}
    \caption{Measurement of magnetic-field fluctuations (at $B_0=5.9~$G), referenced to an AC-line trigger with a variable delay. (a) Reconstructed field noise, as a function of time after an AC-line trigger. The solid line is a fit function $a \cos(\omega_{ac} t + \phi_1) + b \cos(3\omega_{ac}t + \phi_3) + c \cos(5\omega_{ac}t + \phi_5) t$, with $\omega_{ac}=2\pi\times60$~Hz.
    (b) Residual field variation after subtracting the fit function.}
    \label{FIG:ACLine}
\end{figure}

To demonstrate our method, we applied a bias field of 5.9~G using a pair of Helmholtz coils with 10 ppm current stabilization. Figure \ref{FIG:ACLine} shows the results of a typical short-time measurement of the magnetic field along the bias field direction, using an AC-line trigger to start the pulse sequence. The dominant contribution to field fluctuations around $B_0$ is seen to be ambient AC-line noise with an amplitude around $1$~mG, containing the first few harmonics of 60 Hz. From here on, we compensate for this by feeding forward the sign-reversed fit function onto an identical secondary coil of a single winding. The subtraction of the fit results in residual fluctuations up to $\pm0.4$~mG, without apparent phase relationship with the AC-line.

\section{Characterization of performance}

\subsection{Slow Rabi Cycling}
\label{slowrabi}

We characterize the remaining fluctuations further, and in particular determine whether they represent the actual magnetic field in a time interval close to the measurement. For this purpose we implement slow Rabi cycling (at $B_0=9.045~$G) on the maximally magnetic-field sensitive transition $\ket{a}\equiv\ket{1,-1}\leftrightarrow \ket{b}\equiv\ket{2,-2}$, with a differential Zeeman shift of $\gamma=2\pi\times2.1$~kHz/mG. This measurement is performed by varying the coupling time of the oscillation and then  recording the number of atoms in $\ket{b}$. To accommodate the Rabi cycling measurement, we choose a truncated pulse sequence in which the population in $\ket{a}$ is subsequently distributed over five transitions instead of six. We note that this experiment is an example for the mode of operation depicted in Fig.~\ref{FIG:Rb87levels}(a), in which a ``measurement'' (of the Rabi cycling) is followed by a magnetic field ``tag''. Magnetic-field fluctuations will lead to a rapid dephasing of the Rabi oscillation. However, using the field tag, the effect of (slow) magnetic-field fluctuations on the oscillation can be eliminated.

\begin{figure}[h!]
\centering
    \includegraphics[width=1\columnwidth]{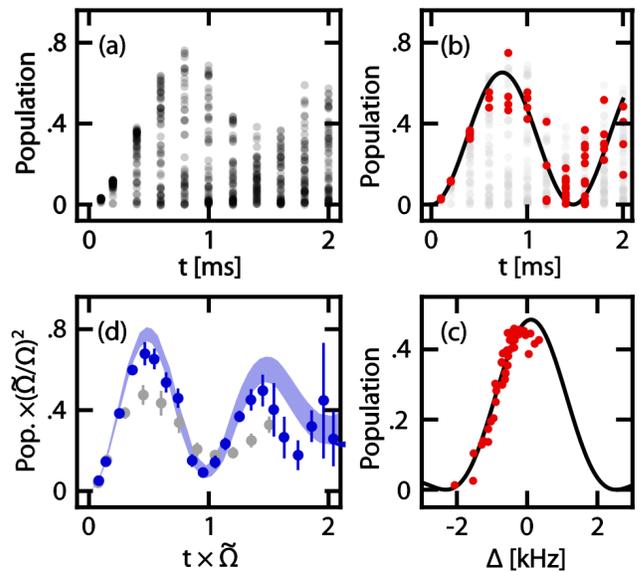}
    \caption{Slow Rabi cycling between $\ket{a}=\ket{1,-1}$ and $\ket{b}=\ket{2,-2}$, with magnetic-field reconstruction based on a 5-pulse sequence. (a) Observed time-dependence of the transferred population after eliminating AC-line fluctuations as demonstrated in Fig. \ref{FIG:ACLine}. The large shot-to-shot scatter is due to residual field fluctuations. (b) Data points post-selected to be within a $100~\mu$G-window. A clear oscillation is recovered, that only dephases after the first cycle. The line is the expected oscillation. (c) Population at a constant time of $400~\mu$s vs. the measured field tag. The solid line is a Rabi resonance fit with the pulse time and Rabi frequency fixed to expectation. (d) Scaled population vs. scaled time. Gray points are original data scaled by average detuning. The shaded line is a simulation, with $B_0$ known to within $55~\mu$G, and $\Omega$ known to within $1~$dB (see text).}
    \label{FIG:RabiCycl}
\end{figure}

For a well-resolved, single-cycle oscillation, the instability of the detuning should not exceed about one tenth of the Rabi frequency. Here we choose $\Omega=2\pi\times0.61(3)~$kHz, at an average detuning of $\delta=2\pi\times0.44(3)~$kHz. 

We see that the raw data resulting from multiple repetitions of the Rabi oscillation experiment has large associated scatter due to the long term drifts and shot-to-shot jitter of the magnetic field. To demonstrate the effect of the field tag, we plot the oscillation both as a function of inferred detuning (at a fixed duration) and time (at a fixed detuning). The results are shown in Fig. \ref{FIG:RabiCycl} (b,c). In addition, we also plot all data, as scaled population $p\tilde{\Omega}^2/\Omega^2$ vs. scaled time $t\tilde{\Omega}$. Clearly, the field tagging leads to a marked improvement of the oscillation contrast.  
 
The next section, \ref{mcsimsec}, will give the details of a simulation of the exact behavior of the field reconstruction. For the given example, and for the parameters of the five-pulse sequence used, we expect the reconstructed fields to scatter around the true magnetic field value with a $55~\mu$G standard-deviation. The simulation and data agree very well, with a slight deviation at late times, potentially due to imperfect cancellation of the AC-line or higher-frequency noise that is uncorrelated with the AC-line.

In our measurements, the high degree of correlation between the transferred population and the detected magnetic field further confirms that the residual fluctuations occur on a scale that is long compared to the duration of the Rabi cycle preceding the field measurement (cf. Fig. \ref{FIG:RabiCycl}). We note that on long time scales, the observed magnetic field drifts are typically on the order of one to several milligauss, over the course of one hour. 

\subsection{Expected theoretical accuracy and operation range} \label{mcsimsec}

For the Rabi oscillation measurements described in section \ref{slowrabi}, the parameters of the magnetometry pulse sequence $i=$ (1, 2, 4, 5, 6) were $\Omega_i/2\pi=$ (2.3, 1.6, 2.6, 2.0, 2.7) kHz, $\tau_i=$ (150, 150, 150, 200, 120) $\mu$s and $\delta_i/2\pi=$ (1.8, 2.8, 2.0, 2.1, 3.4) kHz, which yielded an inferred accuracy of $55~\mu$G. To estimate the ultimate resolution and limits of our magnetometer for optimal parameters (see section \ref{ExpImp}), we perform a Monte-Carlo simulation, using a six-transition sequence. We start with a set of fixed (true) fields $B_{tr}$ drawn from a Gaussian distribution around $B_0$ that are supposed to be reconstructed. The number of atoms transferred in the $i$th pulse at fixed $p_i$ is drawn from a binomial distribution, while the transfer probabilities $p_i$ themselves are subject to uniformly distributed fluctuations of $\tau_i$  ($\pm2~\mu$s), $\Omega_i$ ($\pm1~$dB), $\delta_{i}$ ($\pm2\pi\times7~$Hz) and the instantaneous magnetic field during each individual pulse due to uncanceled residual fluctuations ($\pm100~\mu$G). The Rabi frequencies are $\Omega_i/2\pi=$ (0.9, 1.9, 3.1, 1.2, 1.9, 3.1) kHz and the optimized detunings and pulse areas are $\delta_i=0.82\Omega_i$ and $\tau_i\Omega_i=0.94\pi$ as mentioned earlier. 

\begin{figure}[h!]
\centering
    \includegraphics[width=1\columnwidth]{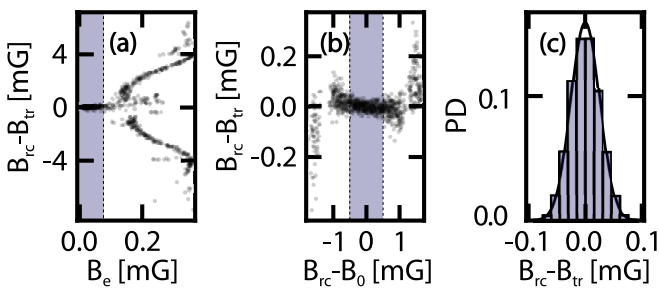}
    \caption{Simulated field reconstruction (over $10^4$ runs).  (a) Reconstruction error vs. fit uncertainty $B_e$, with convergence in the shaded area $B_e<80~\mu$G. (b) Reconstruction error vs. distance of $B_{rc}$ from $B_0$, after discarding fits with  $B_e>80~\mu$G. Proper convergence is obtained in a $\pm500~\mu$G window. (c) Histogram of reconstruction errors for the data in the gray shaded areas (a,b). The solid curve is a Gaussian with a $\sigma$ of $25~\mu$G.}
    \label{FIG:MCSim}
\end{figure}

Results of the simulation are shown in Fig. \ref{FIG:MCSim}. For the optimum pulse parameters, the reconstruction of $B_{tr}$ is accurate to within a standard deviation of $25~\mu$G. A reconstruction is consistently possible within a $\pm500~\mu$G window around $B_0$, if outliers with large fit uncertainties are removed. For larger distances from $B_0$, the default detunings $\delta_{i}$ can be readjusted, or  alternatively, larger Rabi couplings can be used, at the (inversely proportional) expense of the accuracy of the field reconstruction. The results of the simulation confirm that most of the apparent remaining fluctuations in Fig. \ref{FIG:ACLine} are actual fluctuations of the ambient magnetic field (to within the reconstruction uncertainty of $\pm100~\mu$G).

\section{Application: Spectroscopy of state-selective optical lattices}

A number of experimental applications involve the use of homonuclear mixtures of alkali atoms in state-selective optical lattice potentials \cite{RN3002,RN2911,RN2912,RN3203,RN14225,RN19828,RN17365,Lundblad08} , which rely on the existence of a differential Zeeman shift between the states involved. In certain cases, a highly stable separation between a deeply lattice-bound state and a less deeply bound or free state may be desired, such as when the states are subject to coherent coupling \cite{RN2686,RN2665,RN10243}, requiring precise control of both the lattice depth and the magnetic field.

\begin{figure}[h!]
\centering
    \includegraphics[width=1\columnwidth]{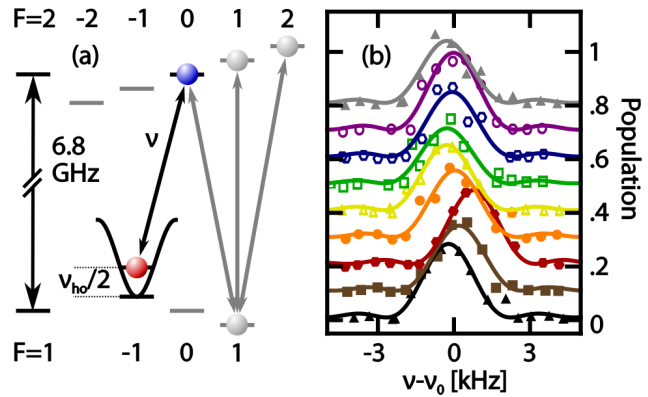}
    \caption{Microwave spectroscopy of a free-to-bound transition in a state-selective optical lattice potential (wavelength 790.10(2)~nm, $\sigma^-$ polarization). (a) Population is transferred from the untrapped state $\ket{b}=\ket{2,0}$ to the confined state $\ket{r}=\ket{1,-1}$. The gray lines indicate the magnetometry sequence following the transfer.  (b) Bound-state population after a 400~$\mu$s long pulse with $\Omega=2\pi\times 450(1)~\rm{Hz}$ and variable detuning, and after accounting for magnetic-field fluctuations. The sequence of spectra was taken at regular intervals over the course of one hour.}
    \label{FIG:SDLSpectr}
\end{figure}

Figure \ref{FIG:SDLSpectr} (a) shows an experimental configuration in which we prepared an ``untrapped'' ensemble of atoms in state $\ket{b}= \ket{2,0}$, coupled to a state $\ket{r}=\ket{1,-1}$  that is confined to the sites of a deep, blue-detuned lattice potential with a zero-point energy shift $h\nu_{ho}/2 = h \times 20(1)$~kHz, generated with circularly polarized light from a titanium-sapphire laser.

To stabilize the magnetic field, we utilize post-selection down to the 100~Hz level based on the magnetic-field tagging described above, using parameters similar to those in Fig. \ref{FIG:MCSim}. The optical intensity $I$ is stabilized to $\sim$1\% using a photodiode and a PID regulation circuit, yielding a transition frequency that should be stable to within about $100~\rm{Hz}$ (since $\omega_{ho}\propto \sqrt{I}$). However, this does not eliminate the possibility of slow drifts of the lattice depth (such as due to temperature induced birefringence or small wavelength changes of the laser) in the course of an experiment, as can be seen in Fig. \ref{FIG:SDLSpectr} (b). To address these issues, the precise resonance condition can now be monitored throughout data taking, using our method. The range of the drift is several hundreds of Hz. We emphasize that the spectroscopic precision necessary for this kind of experiment would not be attainable without first stabilizing the magnetic field.

\section{Conclusions}

In conclusion, we have demonstrated a simple method for in-situ monitoring of magnetic fields in quantum gas experiments with alkali atoms, with a demonstrated accuracy of $55~\mu$G, an inferred accuracy of $25~\mu$G for optimized parameters, and a time resolution of 1~ms. As already seen for the examples above, the magnetometry pulse sequence can be tagged onto experiments that potentially involve several hyperfine states. In principle, the number of transitions used for magnetometry can be reduced down to two, as long as they move differently for a change of the magnetic field (this can be achieved by having two detunings of opposite sign or gyromagnetic ratios of opposite sign). For example, the method could work using only transitions 1 and 4 of Fig. \ref{FIG:Rb87levels} (b). Using a smaller number of transitions generally degrades the accuracy (here by a factor $\sim\sqrt{3}$ when all pulse parameters are left constant, compared to using six transitions), but it increases the measurement bandwidth (here by a factor of 3), which could be an important independent consideration for certain applications.

Thus far, we have only described the use of this method as a scalar magnetometer (in order to be able to ignore fluctuations in perpendicular directions). It should also be possible to access fluctuations of the ambient field in more than one spatial direction, if the bias field is rotated during the magnetometry pulse sequence (with two transitions used per direction). This can become important if one wants to use this method for stable field post-selection at low fields. 

Finally, for comparison to other magnetometry techniques, a sensitivity may be specified as \cite{RN21456}
$
\eta = \Delta B_{min}\sqrt T
$,
i.e. as the minimum detectable change in field $\Delta B_{min}=2\sqrt{\ln2}\sigma\sim$60~$\mu$G multiplied by the square root of the cycle time. Since typical field fluctuations in laboratories usually stem from AC-mains or are very low frequency (such as fluctuations of Earth's magnetic field), synchronizing the experiment to the AC-line can yield one measurement in the effective integration time of 1~ms. In this case, an effective sensitivity $\eta\sim300~\rm{pT/\sqrt{Hz}}$ (in a measurement volume of 10~$\mu$m$^3$) can be reached.

\section{Acknowledgements}
We thank M. G. Cohen for discussions and a critical reading of the manuscript. This work was supported by NSF PHY-1205894 and PHY-1607633. M. S. was supported from a GAANN fellowship by the DoEd. A. P. acknowledges partial support from EPSOL-SENESCYT.

\section*{References}
\bibliographystyle{apsrev}
\bibliography{MagRefs}

\begin{thebibliography}{26}
\expandafter\ifx\csname natexlab\endcsname\relax\def\natexlab#1{#1}\fi
\expandafter\ifx\csname bibnamefont\endcsname\relax
  \def\bibnamefont#1{#1}\fi
\expandafter\ifx\csname bibfnamefont\endcsname\relax
  \def\bibfnamefont#1{#1}\fi
\expandafter\ifx\csname citenamefont\endcsname\relax
  \def\citenamefont#1{#1}\fi
\expandafter\ifx\csname url\endcsname\relax
  \def\url#1{\texttt{#1}}\fi
\expandafter\ifx\csname urlprefix\endcsname\relax\def\urlprefix{URL }\fi
\providecommand{\bibinfo}[2]{#2}
\providecommand{\eprint}[2][]{\url{#2}}

\bibitem[{\citenamefont{Pethick and Smith}(2008)}]{RN28382}
\bibinfo{author}{\bibfnamefont{C.~J.} \bibnamefont{Pethick}} \bibnamefont{and}
  \bibinfo{author}{\bibfnamefont{H.}~\bibnamefont{Smith}},
  \emph{\bibinfo{title}{Bose-Einstein Condensation in Dilute Gases}}
  (\bibinfo{publisher}{Cambridge University Press}, \bibinfo{year}{2008}),
  \bibinfo{edition}{2nd} ed.

\bibitem[{\citenamefont{Stamper-Kurn and Ueda}(2013)}]{RN12416}
\bibinfo{author}{\bibfnamefont{D.~M.} \bibnamefont{Stamper-Kurn}}
  \bibnamefont{and} \bibinfo{author}{\bibfnamefont{M.}~\bibnamefont{Ueda}},
  \bibinfo{journal}{Reviews of Modern Physics} \textbf{\bibinfo{volume}{85}},
  \bibinfo{pages}{1191} (\bibinfo{year}{2013}),
  \urlprefix\url{http://link.aps.org/doi/10.1103/RevModPhys.85.1191}.

\bibitem[{\citenamefont{Deutsch and Jessen}(1998)}]{RN3002}
\bibinfo{author}{\bibfnamefont{I.~H.} \bibnamefont{Deutsch}} \bibnamefont{and}
  \bibinfo{author}{\bibfnamefont{P.~S.} \bibnamefont{Jessen}},
  \bibinfo{journal}{Physical Review A} \textbf{\bibinfo{volume}{57}},
  \bibinfo{pages}{1972} (\bibinfo{year}{1998}),
  \urlprefix\url{http://link.aps.org/doi/10.1103/PhysRevA.57.1972}.

\bibitem[{\citenamefont{Pertot et~al.}(2010)\citenamefont{Pertot, Gadway, and
  Schneble}}]{RN2911}
\bibinfo{author}{\bibfnamefont{D.}~\bibnamefont{Pertot}},
  \bibinfo{author}{\bibfnamefont{B.}~\bibnamefont{Gadway}}, \bibnamefont{and}
  \bibinfo{author}{\bibfnamefont{D.}~\bibnamefont{Schneble}},
  \bibinfo{journal}{Physical Review Letters} \textbf{\bibinfo{volume}{104}}
  (\bibinfo{year}{2010}), ISSN \bibinfo{issn}{0031-9007 1079-7114}.

\bibitem[{\citenamefont{Gadway et~al.}(2010)\citenamefont{Gadway, Pertot,
  Reimann, and Schneble}}]{RN2912}
\bibinfo{author}{\bibfnamefont{B.}~\bibnamefont{Gadway}},
  \bibinfo{author}{\bibfnamefont{D.}~\bibnamefont{Pertot}},
  \bibinfo{author}{\bibfnamefont{R.}~\bibnamefont{Reimann}}, \bibnamefont{and}
  \bibinfo{author}{\bibfnamefont{D.}~\bibnamefont{Schneble}},
  \bibinfo{journal}{Physical Review Letters} \textbf{\bibinfo{volume}{105}}
  (\bibinfo{year}{2010}), ISSN \bibinfo{issn}{0031-9007 1079-7114}.

\bibitem[{\citenamefont{Gadway et~al.}(2011)\citenamefont{Gadway, Pertot,
  Reeves, Vogt, and Schneble}}]{RN3203}
\bibinfo{author}{\bibfnamefont{B.}~\bibnamefont{Gadway}},
  \bibinfo{author}{\bibfnamefont{D.}~\bibnamefont{Pertot}},
  \bibinfo{author}{\bibfnamefont{J.}~\bibnamefont{Reeves}},
  \bibinfo{author}{\bibfnamefont{M.}~\bibnamefont{Vogt}}, \bibnamefont{and}
  \bibinfo{author}{\bibfnamefont{D.}~\bibnamefont{Schneble}},
  \bibinfo{journal}{Physical Review Letters} \textbf{\bibinfo{volume}{107}},
  \bibinfo{pages}{145306} (\bibinfo{year}{2011}), ISSN \bibinfo{issn}{1079-7114
  (Electronic) 0031-9007 (Linking)},
  \urlprefix\url{http://www.ncbi.nlm.nih.gov/pubmed/22107210}.

\bibitem[{\citenamefont{Gadway et~al.}(2012)\citenamefont{Gadway, Pertot,
  Reeves, and Schneble}}]{RN14225}
\bibinfo{author}{\bibfnamefont{B.}~\bibnamefont{Gadway}},
  \bibinfo{author}{\bibfnamefont{D.}~\bibnamefont{Pertot}},
  \bibinfo{author}{\bibfnamefont{J.}~\bibnamefont{Reeves}}, \bibnamefont{and}
  \bibinfo{author}{\bibfnamefont{D.}~\bibnamefont{Schneble}},
  \bibinfo{journal}{Nature Physics} \textbf{\bibinfo{volume}{8}},
  \bibinfo{pages}{544} (\bibinfo{year}{2012}), ISSN \bibinfo{issn}{1745-2473},
  \urlprefix\url{<Go to ISI>://WOS:000305970400014}.

\bibitem[{\citenamefont{Reeves et~al.}(2015)\citenamefont{Reeves, Krinner,
  Stewart, Pazmi\~no, and Schneble}}]{RN19828}
\bibinfo{author}{\bibfnamefont{J.}~\bibnamefont{Reeves}},
  \bibinfo{author}{\bibfnamefont{L.}~\bibnamefont{Krinner}},
  \bibinfo{author}{\bibfnamefont{M.}~\bibnamefont{Stewart}},
  \bibinfo{author}{\bibfnamefont{A.}~\bibnamefont{Pazmi\~no}},
  \bibnamefont{and} \bibinfo{author}{\bibfnamefont{D.}~\bibnamefont{Schneble}},
  \bibinfo{journal}{Physical Review A} \textbf{\bibinfo{volume}{92}},
  \bibinfo{pages}{023628} (\bibinfo{year}{2015}),
  \urlprefix\url{http://link.aps.org/doi/10.1103/PhysRevA.92.023628}.

\bibitem[{\citenamefont{McKay and DeMarco}(2010)}]{RN17365}
\bibinfo{author}{\bibfnamefont{D.}~\bibnamefont{McKay}} \bibnamefont{and}
  \bibinfo{author}{\bibfnamefont{B.}~\bibnamefont{DeMarco}},
  \bibinfo{journal}{New Journal of Physics} \textbf{\bibinfo{volume}{12}}
  (\bibinfo{year}{2010}), ISSN \bibinfo{issn}{1367-2630}, \urlprefix\url{<Go to
  ISI>://WOS:000278634100008}.

\bibitem[{\citenamefont{McKay and DeMarco}(2011)}]{RN3055}
\bibinfo{author}{\bibfnamefont{D.~C.} \bibnamefont{McKay}} \bibnamefont{and}
  \bibinfo{author}{\bibfnamefont{B.}~\bibnamefont{DeMarco}},
  \bibinfo{journal}{Reports on Progress in Physics}
  \textbf{\bibinfo{volume}{74}}, \bibinfo{pages}{054401}
  (\bibinfo{year}{2011}), ISSN \bibinfo{issn}{0034-4885 1361-6633},
  \urlprefix\url{http://arxiv.org/pdf/1010.0198v2}.

\bibitem[{\citenamefont{Kominis et~al.}(2003)\citenamefont{Kominis, Kornack,
  Allred, and Romalis}}]{RN28384}
\bibinfo{author}{\bibfnamefont{I.~K.} \bibnamefont{Kominis}},
  \bibinfo{author}{\bibfnamefont{T.~W.} \bibnamefont{Kornack}},
  \bibinfo{author}{\bibfnamefont{J.~C.} \bibnamefont{Allred}},
  \bibnamefont{and} \bibinfo{author}{\bibfnamefont{M.~V.}
  \bibnamefont{Romalis}}, \bibinfo{journal}{Nature}
  \textbf{\bibinfo{volume}{422}}, \bibinfo{pages}{596} (\bibinfo{year}{2003}),
  ISSN \bibinfo{issn}{0028-0836},
  \urlprefix\url{http://dx.doi.org/10.1038/nature01484}.

\bibitem[{\citenamefont{Wasilewski et~al.}(2010)\citenamefont{Wasilewski,
  Jensen, Krauter, Renema, Balabas, and Polzik}}]{RN28385}
\bibinfo{author}{\bibfnamefont{W.}~\bibnamefont{Wasilewski}},
  \bibinfo{author}{\bibfnamefont{K.}~\bibnamefont{Jensen}},
  \bibinfo{author}{\bibfnamefont{H.}~\bibnamefont{Krauter}},
  \bibinfo{author}{\bibfnamefont{J.~J.} \bibnamefont{Renema}},
  \bibinfo{author}{\bibfnamefont{M.~V.} \bibnamefont{Balabas}},
  \bibnamefont{and} \bibinfo{author}{\bibfnamefont{E.~S.}
  \bibnamefont{Polzik}}, \bibinfo{journal}{Physical Review Letters}
  \textbf{\bibinfo{volume}{104}}, \bibinfo{pages}{133601}
  (\bibinfo{year}{2010}),
  \urlprefix\url{https://link.aps.org/doi/10.1103/PhysRevLett.104.133601}.

\bibitem[{\citenamefont{Wildermuth et~al.}(2005)\citenamefont{Wildermuth,
  Hofferberth, Lesanovsky, Haller, Andersson, Groth, Bar-Joseph, Kruger, and
  Schmiedmayer}}]{RN28387}
\bibinfo{author}{\bibfnamefont{S.}~\bibnamefont{Wildermuth}},
  \bibinfo{author}{\bibfnamefont{S.}~\bibnamefont{Hofferberth}},
  \bibinfo{author}{\bibfnamefont{I.}~\bibnamefont{Lesanovsky}},
  \bibinfo{author}{\bibfnamefont{E.}~\bibnamefont{Haller}},
  \bibinfo{author}{\bibfnamefont{L.~M.} \bibnamefont{Andersson}},
  \bibinfo{author}{\bibfnamefont{S.}~\bibnamefont{Groth}},
  \bibinfo{author}{\bibfnamefont{I.}~\bibnamefont{Bar-Joseph}},
  \bibinfo{author}{\bibfnamefont{P.}~\bibnamefont{Kruger}}, \bibnamefont{and}
  \bibinfo{author}{\bibfnamefont{J.}~\bibnamefont{Schmiedmayer}},
  \bibinfo{journal}{Nature} \textbf{\bibinfo{volume}{435}},
  \bibinfo{pages}{440} (\bibinfo{year}{2005}), ISSN \bibinfo{issn}{0028-0836},
  \urlprefix\url{http://dx.doi.org/10.1038/435440a}.

\bibitem[{\citenamefont{Vengalattore et~al.}(2007)\citenamefont{Vengalattore,
  Higbie, Leslie, Guzman, Sadler, and Stamper-Kurn}}]{RN2693}
\bibinfo{author}{\bibfnamefont{M.}~\bibnamefont{Vengalattore}},
  \bibinfo{author}{\bibfnamefont{J.~M.} \bibnamefont{Higbie}},
  \bibinfo{author}{\bibfnamefont{S.~R.} \bibnamefont{Leslie}},
  \bibinfo{author}{\bibfnamefont{J.}~\bibnamefont{Guzman}},
  \bibinfo{author}{\bibfnamefont{L.~E.} \bibnamefont{Sadler}},
  \bibnamefont{and} \bibinfo{author}{\bibfnamefont{D.~M.}
  \bibnamefont{Stamper-Kurn}}, \bibinfo{journal}{Physical Review Letters}
  \textbf{\bibinfo{volume}{98}} (\bibinfo{year}{2007}), ISSN
  \bibinfo{issn}{0031-9007 1079-7114}.

\bibitem[{\citenamefont{Yang et~al.}(2017)\citenamefont{Yang, Koll\'ar, Taylor,
  Turner, and Lev}}]{RN28386}
\bibinfo{author}{\bibfnamefont{F.}~\bibnamefont{Yang}},
  \bibinfo{author}{\bibfnamefont{A.~J.} \bibnamefont{Koll\'ar}},
  \bibinfo{author}{\bibfnamefont{S.~F.} \bibnamefont{Taylor}},
  \bibinfo{author}{\bibfnamefont{R.~W.} \bibnamefont{Turner}},
  \bibnamefont{and} \bibinfo{author}{\bibfnamefont{B.~L.} \bibnamefont{Lev}},
  \bibinfo{journal}{Physical Review Applied} \textbf{\bibinfo{volume}{7}},
  \bibinfo{pages}{034026} (\bibinfo{year}{2017}),
  \urlprefix\url{https://link.aps.org/doi/10.1103/PhysRevApplied.7.034026}.

\bibitem[{\citenamefont{Muessel et~al.}(2014)\citenamefont{Muessel, Strobel,
  Linnemann, Hume, and Oberthaler}}]{RN4624}
\bibinfo{author}{\bibfnamefont{W.}~\bibnamefont{Muessel}},
  \bibinfo{author}{\bibfnamefont{H.}~\bibnamefont{Strobel}},
  \bibinfo{author}{\bibfnamefont{D.}~\bibnamefont{Linnemann}},
  \bibinfo{author}{\bibfnamefont{D.~B.} \bibnamefont{Hume}}, \bibnamefont{and}
  \bibinfo{author}{\bibfnamefont{M.~K.} \bibnamefont{Oberthaler}},
  \bibinfo{journal}{Physical Review Letters} \textbf{\bibinfo{volume}{113}},
  \bibinfo{pages}{103004} (\bibinfo{year}{2014}), ISSN \bibinfo{issn}{1079-7114
  (Electronic) 0031-9007 (Linking)},
  \urlprefix\url{http://www.ncbi.nlm.nih.gov/pubmed/25238356}.

\bibitem[{\citenamefont{B\"ohi et~al.}(2009)\citenamefont{B\"ohi, Riedel,
  Hoffrogge, Reichel, H\"ansch, and Treutlein}}]{RN2804}
\bibinfo{author}{\bibfnamefont{P.}~\bibnamefont{B\"ohi}},
  \bibinfo{author}{\bibfnamefont{M.~F.} \bibnamefont{Riedel}},
  \bibinfo{author}{\bibfnamefont{J.}~\bibnamefont{Hoffrogge}},
  \bibinfo{author}{\bibfnamefont{J.}~\bibnamefont{Reichel}},
  \bibinfo{author}{\bibfnamefont{T.~W.} \bibnamefont{H\"ansch}},
  \bibnamefont{and}
  \bibinfo{author}{\bibfnamefont{P.}~\bibnamefont{Treutlein}},
  \bibinfo{journal}{Nature Physics} \textbf{\bibinfo{volume}{5}},
  \bibinfo{pages}{592} (\bibinfo{year}{2009}), ISSN \bibinfo{issn}{1745-2473
  1745-2481}.

\bibitem[{\citenamefont{Gro{\ss}}(2010)}]{RN4429}
\bibinfo{author}{\bibfnamefont{C.}~\bibnamefont{Gro{\ss}}},
  \bibinfo{journal}{Ph.D. thesis, Universit\"at Heidelberg}
  (\bibinfo{year}{2010}).

\bibitem[{Note1()}]{Note1}
Note1, \bibinfo{note}{for $\delta B \ll B_0$, fluctuations perpendicular to z
  can be neglected, since they are quadratically suppressed.}

\bibitem[{\citenamefont{Steck}(revision 2.1.5, 13 January 2015)}]{RN3163}
\bibinfo{author}{\bibfnamefont{D.~A.} \bibnamefont{Steck}},
  \bibinfo{journal}{Rubidium 87 D Line Data, available online at
  http://steck.us/alkalidata}  (\bibinfo{year}{revision 2.1.5, 13 January
  2015}).

\bibitem[{\citenamefont{Pertot et~al.}(2009)\citenamefont{Pertot, Greif,
  Albert, Gadway, and Schneble}}]{Pertot-09-Machine}
\bibinfo{author}{\bibfnamefont{D.}~\bibnamefont{Pertot}},
  \bibinfo{author}{\bibfnamefont{D.}~\bibnamefont{Greif}},
  \bibinfo{author}{\bibfnamefont{S.}~\bibnamefont{Albert}},
  \bibinfo{author}{\bibfnamefont{B.}~\bibnamefont{Gadway}}, \bibnamefont{and}
  \bibinfo{author}{\bibfnamefont{D.}~\bibnamefont{Schneble}},
  \bibinfo{journal}{J. Phys. B} \textbf{\bibinfo{volume}{42}},
  \bibinfo{pages}{215305} (\bibinfo{year}{2009}),
  \urlprefix\url{http://stacks.iop.org/0953-4075/42/i=21/a=215305}.

\bibitem[{\citenamefont{Lundblad et~al.}(2008)\citenamefont{Lundblad, Lee,
  Spielman, Brown, Phillips, and Porto}}]{Lundblad08}
\bibinfo{author}{\bibfnamefont{N.}~\bibnamefont{Lundblad}},
  \bibinfo{author}{\bibfnamefont{P.~J.} \bibnamefont{Lee}},
  \bibinfo{author}{\bibfnamefont{I.~B.} \bibnamefont{Spielman}},
  \bibinfo{author}{\bibfnamefont{B.~L.} \bibnamefont{Brown}},
  \bibinfo{author}{\bibfnamefont{W.~D.} \bibnamefont{Phillips}},
  \bibnamefont{and} \bibinfo{author}{\bibfnamefont{J.~V.} \bibnamefont{Porto}},
  \bibinfo{journal}{Phys. Rev. Lett.} \textbf{\bibinfo{volume}{100}},
  \bibinfo{pages}{150401} (\bibinfo{year}{2008}),
  \urlprefix\url{https://link.aps.org/doi/10.1103/PhysRevLett.100.150401}.

\bibitem[{\citenamefont{Recati et~al.}(2005)\citenamefont{Recati, Fedichev,
  Zwerger, von Delft, and Zoller}}]{RN2686}
\bibinfo{author}{\bibfnamefont{A.}~\bibnamefont{Recati}},
  \bibinfo{author}{\bibfnamefont{P.~O.} \bibnamefont{Fedichev}},
  \bibinfo{author}{\bibfnamefont{W.}~\bibnamefont{Zwerger}},
  \bibinfo{author}{\bibfnamefont{J.}~\bibnamefont{von Delft}},
  \bibnamefont{and} \bibinfo{author}{\bibfnamefont{P.}~\bibnamefont{Zoller}},
  \bibinfo{journal}{Physical Review Letters} \textbf{\bibinfo{volume}{94}}
  (\bibinfo{year}{2005}), ISSN \bibinfo{issn}{0031-9007 1079-7114}.

\bibitem[{\citenamefont{Orth et~al.}(2008)\citenamefont{Orth, Stanic, and
  Le~Hur}}]{RN2665}
\bibinfo{author}{\bibfnamefont{P.~P.} \bibnamefont{Orth}},
  \bibinfo{author}{\bibfnamefont{I.}~\bibnamefont{Stanic}}, \bibnamefont{and}
  \bibinfo{author}{\bibfnamefont{K.}~\bibnamefont{Le~Hur}},
  \bibinfo{journal}{Physical Review A} \textbf{\bibinfo{volume}{77}}
  (\bibinfo{year}{2008}), ISSN \bibinfo{issn}{1050-2947 1094-1622}.

\bibitem[{\citenamefont{de~Vega et~al.}(2008)\citenamefont{de~Vega, Porras, and
  Ignacio~Cirac}}]{RN10243}
\bibinfo{author}{\bibfnamefont{I.}~\bibnamefont{de~Vega}},
  \bibinfo{author}{\bibfnamefont{D.}~\bibnamefont{Porras}}, \bibnamefont{and}
  \bibinfo{author}{\bibfnamefont{J.}~\bibnamefont{Ignacio~Cirac}},
  \bibinfo{journal}{Physical Review Letters} \textbf{\bibinfo{volume}{101}},
  \bibinfo{pages}{260404} (\bibinfo{year}{2008}),
  \urlprefix\url{http://link.aps.org/doi/10.1103/PhysRevLett.101.260404}.

\bibitem[{\citenamefont{Taylor et~al.}(2008)\citenamefont{Taylor, Cappellaro,
  Childress, Jiang, Budker, Hemmer, Yacoby, Walsworth, and Lukin}}]{RN21456}
\bibinfo{author}{\bibfnamefont{J.~M.} \bibnamefont{Taylor}},
  \bibinfo{author}{\bibfnamefont{P.}~\bibnamefont{Cappellaro}},
  \bibinfo{author}{\bibfnamefont{L.}~\bibnamefont{Childress}},
  \bibinfo{author}{\bibfnamefont{L.}~\bibnamefont{Jiang}},
  \bibinfo{author}{\bibfnamefont{D.}~\bibnamefont{Budker}},
  \bibinfo{author}{\bibfnamefont{P.~R.} \bibnamefont{Hemmer}},
  \bibinfo{author}{\bibfnamefont{A.}~\bibnamefont{Yacoby}},
  \bibinfo{author}{\bibfnamefont{R.}~\bibnamefont{Walsworth}},
  \bibnamefont{and} \bibinfo{author}{\bibfnamefont{M.~D.} \bibnamefont{Lukin}},
  \bibinfo{journal}{Nat Phys} \textbf{\bibinfo{volume}{4}},
  \bibinfo{pages}{810} (\bibinfo{year}{2008}), ISSN \bibinfo{issn}{1745-2473},
  \urlprefix\url{http://dx.doi.org/10.1038/nphys1075}.

\end{thebibliography}
\end{document}